\documentclass[aps,prb,reprint,groupedaddress,amsmath,amssymb]{revtex4-2}
\usepackage[T1]{fontenc}
\usepackage[utf8]{inputenc}
\usepackage{graphicx}
\usepackage{upgreek}
\usepackage{hyperref}

\begin{document}
\title{Proximity Transient Absorption in van der Waals Heterostructures}
\author{Neema Rafizadeh}
\author{Ting Zheng}
\author{Hui Zhao}
\email{huizhao@ku.edu}
\affiliation{Department of Physics and Astronomy, The University of Kansas, Lawrence, Kansas 66045, United States}
\begin{abstract}

Transient absorption spectroscopy is a widely used technique for investigating ultrafast electronic dynamics, where transient optical signals are generally interpreted as arising from excitations within the material hosting the probed optical transition. A distinct form of transient absorption spectroscopy, termed proximity transient absorption (PTA), is demonstrated, in which the probed optical transition and the monitored photoexcited carriers reside in different materials. Using graphene/hBN/WS$_2$ heterostructures, photoexcited carriers in graphene induce a strong transient optical response at the excitonic resonance of a WS$_2$ monolayer through Coulomb screening, even when the two layers remain electronically isolated by an atomically thin hBN spacer. The resulting signal follows the carrier dynamics of graphene rather than those of WS$_2$, demonstrating that the WS$_2$ exciton acts as an optical transducer of electronic processes in a neighboring material. The signal decreases systematically with increasing hBN thickness, confirming its proximity origin while remaining substantial for separations approaching 2 nm. PTA provides a noncontact approach for optically modulating one material using photoexcited carriers in another and enables probing of carrier dynamics in materials that are otherwise difficult to investigate directly. More broadly, PTA establishes a new modality of ultrafast spectroscopy and reveals that transient optical signals in van der Waals heterostructures can originate from excitations residing in spatially separated layers.

\end{abstract}
\maketitle

\section{Introduction}
Light absorption is a fundamental property of atoms, molecules, and condensed matter systems. Together with reflection and transmission, it is determined by the dielectric function, or equivalently the complex index of refraction, of the material. Although these quantities are often regarded as intrinsic material properties, they can be modified by changes in the state of the system, such as variations in electron population, temperature, pressure, electric field, magnetic field, or crystal structure. Consequently, when a material is subjected to a transient perturbation, its optical absorption also changes transiently. Monitoring this change, known as transient absorption, and its subsequent recovery provides a powerful means of tracking the evolution of the underlying physical processes.

Transient absorption spectroscopy is one of the most widely used techniques for measuring such dynamics.\cite{jpca1045660,bookshah} In a typical experiment, a pump pulse or other external perturbation drives the system away from equilibrium, while a time-delayed probe pulse monitors the resulting change in optical absorption by measuring reflectance or transmittance. By varying the time delay between the perturbation and probe, the temporal evolution of the transient state can be followed with high time resolution, often reaching the femtosecond regime. 

Although transient absorption spectroscopy can probe a broad range of phenomena, including thermal, structural, magnetic, and electronic dynamics, one of its most widespread applications is the investigation of photoinduced electronic processes. In atomic and molecular systems, the interpretation of transient absorption due to photoexcitation is often comparatively straightforward because the signal can frequently be interpreted in terms of population changes among a relatively small number of well-defined electronic states.\cite{n419803,n466739,jpca1045660} In condensed matter systems, however, many-body interactions substantially complicate this picture. Through electron--electron and electron--hole interactions, photoexcited carriers occupying one set of states can modify optical transitions involving different states. \cite{bookshah} For example, in semiconductors and their quantum wells, photoexcited electron-hole pairs, occupying continuum states in the conduction and valence bands, respectively, can produce large transient absorption at excitonic resonances through Coulomb screening, phase-space filling, and bandgap renormalization.\cite{b326601} These effects are particularly pronounced in two-dimensional semiconductors, \cite{nphoto9466} where reduced dielectric screening enhances Coulomb interactions.\cite{l113026803,l113076802} Consequently, the relationship between transient absorption and carrier populations is often indirect and may involve states far removed from those being optically probed. Nevertheless, in most cases, the transient optical response of a given material is still assumed to reflect the dynamics of carriers, excitons, or other excitations residing within that same material.

Two-dimensional (2D) materials based on van der Waals (vdW) crystals provide a unique platform for extending transient absorption spectroscopy beyond its conventional form. Unlike bulk materials, vdW heterostructures allow distinct atomically thin materials to be brought into atomic-scale proximity while retaining their individual electronic and optical properties.\cite{s335947,n499419} Consequently, the optical response of one layer can be influenced by the presence of a neighboring layer even when the two materials remain electronically distinct. Such proximity effects arise from Coulomb interactions mediated across the atomically thin interface, providing new opportunities for manipulating and sensing optical excitations across interfaces.

Transient absorption spectroscopy has been widely used to investigate photoexcitation dynamics in 2D materials, including carrier thermalization and energy relaxation,\cite{acsnano810931,nc813906} exciton formation,\cite{nanoscale811681,b102035414}   recombination,\cite{acsnano71072,acsnano811147,nl158204,nm14889,b89125427,nl145625} transport,\cite{acsnano96436,acsnano82970} and spin/valley relaxation,\cite{acsnano711087,nc68963,np11830,s3551066} as well as coherent excitonic processes.\cite{nm14290,s3461205} The emergence of vdW heterostructures has further expanded the application of transient absorption spectroscopy. Since the first ultrafast studies of vdW heterostructures in 2014, transient absorption spectroscopy has been extensively applied to van der Waals heterostructures, including semiconducting TMD/TMD heterostructures,\cite{nn9682,acsnano812717,nanoscale717523,acsnano106612,acsnano1112020,2dm4015033,nl173591,jmr351417,np151140,b99195438,nl2210140,acsnano18931} graphene/TMD heterostructures,\cite{sa4e1700324,nc113876,nanoscale122498,jpcl135123,acsp115170,acsnano1516760,afm322201123} heterostructures formed between semiconducting TMDs and other 2D materials,\cite{acsami1253475,nanoscale914533,nh4236,small202304010} multilayer heterostructures,\cite{nl171623,jpcl95970,acsami1312425,apl118253106,acsnano173939} and moir\'e or correlated heterostructures.\cite{nm19617,nm24527} In these studies, a transient signal observed at the optical resonance of a given layer is generally interpreted as reflecting the dynamics of carriers, excitons, or other excitations residing within that same layer. The unique geometry of vdW heterostructures, however, enables a fundamentally different possibility: can transient absorption at the optical resonance of one material originate primarily from electronic processes occurring in a nearby material while the two materials remain electronically distinct? Realizing such a phenomenon would establish a new form of transient absorption spectroscopy, enabling remote sensing and manipulation of electronic processes through proximity interactions.

Here we investigate this concept by demonstrating a distinct form of transient absorption spectroscopy, which we term proximity transient absorption (PTA), in which the optical transition being probed and the photoexcited carriers being monitored reside in different materials. Using graphene/hBN/WS$_2$ heterostructures as a model system, we show that carriers generated in graphene (Gr) induce a strong transient optical response at the excitonic resonance of a nearby WS$_2$ monolayer (1L) through Coulomb screening, even when the two layers remain electronically isolated by an atomically thin hexagonal boron nitride (hBN) spacer. Beyond its fundamental interest, PTA offers a noninvasive means of modulating the optical properties of a semiconductor without directly injecting carriers into the active material and provides a new approach for probing carrier dynamics in materials that are difficult to study using conventional transient absorption spectroscopy, such as indirect-gap semiconductors, materials with weak optical oscillator strength, or systems whose relevant optical transitions lie outside the spectral range accessible to available laser sources. More broadly, our results demonstrate that a transient signal observed at the optical resonance of a given layer in a heterostructure does not necessarily originate from excitations within that layer, but can instead be generated by electronic processes occurring in a nearby material through proximity interactions. Finally, by comparing heterostructures with and without an hBN spacer, we obtain additional evidence for the emergence of coupled electronic states in directly contacted Gr/WS$_2$ heterostructures. Together, these findings establish PTA as a new tool for ultrafast spectroscopy and reveal new opportunities enabled by proximity effects in atomically thin materials.

\section{Results and Discussion}

\subsection{Sample Design, Fabrication, and Characterization}

\begin{figure}[t]
\centerline{\includegraphics[width=\columnwidth]{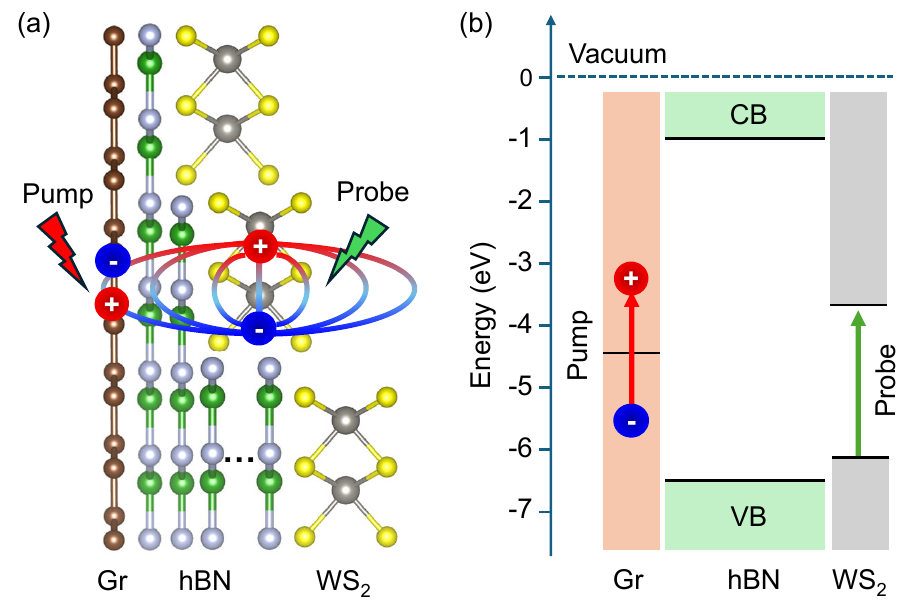}}
\caption{Schematic illustration of proximity transient absorption (PTA). (a) Graphene and WS$_2$ monolayers are separated by a thin hexagonal boron nitride (hBN) layer. A pump pulse generates electron--hole pairs in graphene. The resulting carriers induce Coulomb screening across the hBN spacer, modifying the excitonic resonance of the nearby WS$_2$ layer. The transient change in the WS$_2$ optical response is detected by a probe pulse resonant with the exciton transition. (b) Schematic band alignment and optical transitions in the Gr/hBN/WS$_2$ heterostructure.} 
\label{Fig:design}
\end{figure}

Figure~\ref{Fig:design}(a) illustrates the heterostructure used to investigate PTA, in which a 1L graphene and a 1L WS$_2$ are separated by an hBN spacer with various thicknesses. The corresponding band alignment is shown schematically in Figure~\ref{Fig:design}(b). We adopt electron affinities of 0.89 and 3.76 eV, ionization potentials of 6.60 and 6.10 eV for hBN and WS$_2$, respectively, and a graphene Dirac-point energy of 4.34 eV below vacuum, all obtained from the same HSE06 calculation.\cite{jpcl1311581} Although the exact values may depend on thickness and dielectric environment, they clearly establish that hBN presents a large barrier for both electrons and holes, thereby suppressing charge transfer and maintaining electronic isolation between the graphene and WS$_2$ layers. When electrons and holes are photoexcited in the graphene layer, they modify the excitonic resonances of the WS$_2$ 1L through carrier-induced Coulomb screening across the hBN spacer. As a result, the optical absorption of WS$_2$ is altered transiently and can be detected by a probe pulse tuned near the excitonic resonance of WS$_2$.

\begin{figure}[t]
\centerline{\includegraphics[width=\columnwidth]{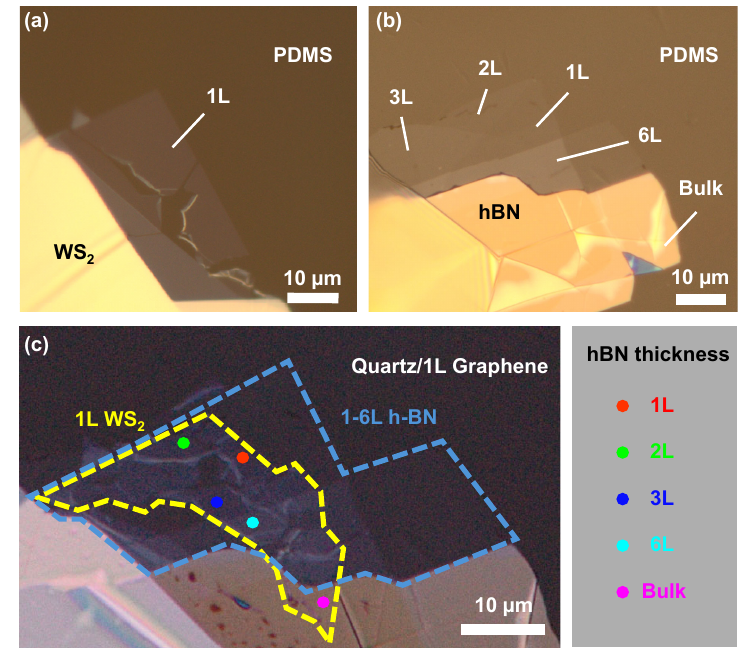}}
\caption{Optical microscope images of the samples. (a) A WS$_2$ flake exfoliated onto a polydimethylsiloxane (PDMS) substrate, containing a large monolayer (1L) region. (b) An hBN flake exfoliated onto a PDMS substrate, containing regions of different thicknesses, as labeled, determined from their optical contrast. (c) The Gr/hBN/WS$_2$ heterostructure fabricated by first transferring the hBN flake onto CVD monolayer graphene on a quartz substrate, followed by transfer of the 1L WS$_2$ flake. Colored dots indicate the regions with different hBN thicknesses and location and size of laser spots in optical measurements.} 
\label{Fig:sample}
\end{figure}

To fabricate these samples, a 1L WS$_2$ flake and a hBN flake with various thicknesses were mechanically exfoliated onto transparent polydimethylsiloxane (PDMS) substrates, as shown in Figure~\ref{Fig:sample}(a) and \ref{Fig:sample}(b), respectively. The thickness of the flakes was determined from their optical contrast on PDMS (see Figure S1). The hBN flake was then transferred onto a graphene film grown by chemical vapor deposition (CVD) on a quartz substrate (acquired from ACS Materials). Next, the 1L WS$_2$ flake was transferred onto the graphene/hBN structure, creating a series of Gr/hBN/WS$_2$ heterostructures with different hBN thicknesses, as shown in Figure~\ref{Fig:sample}(c).

The samples were characterized by photoluminescence (PL) spectroscopy at room temperature and under ambient conditions. The samples were excited by a 3.06~eV continuous-wave laser beam with a power of 1~$\upmu$W. The laser beam was focused onto the sample using a microscope objective with a numerical aperture of 0.7, allowing it to be positioned safely within each region of the sample, as indicated by the colored dots in Figure~\ref{Fig:sample}(c). PL spectra were acquired with an integration time of 10~s.

\begin{figure}[t]
\centerline{\includegraphics[width=\columnwidth]{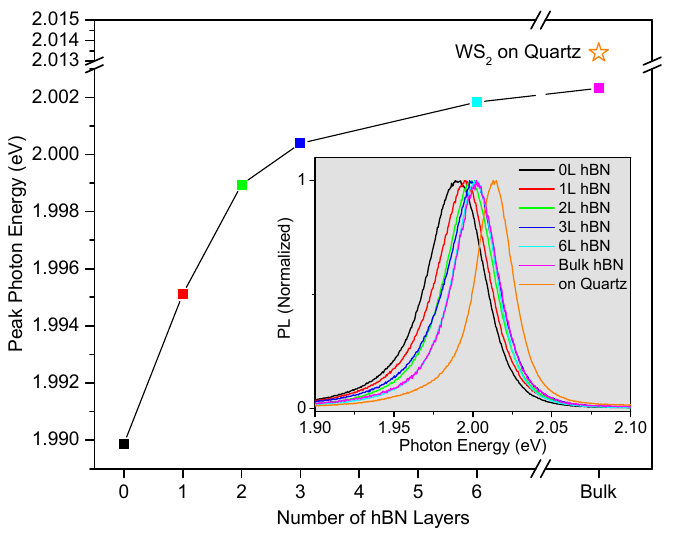}}
\caption{Photoluminescence (PL) spectroscopy. The inset shows normalized PL spectra measured from different regions of the heterostructure sample and from a monolayer WS$_2$ sample on quartz. The main panel shows the WS$_2$ PL peak energy as a function of the hBN layer number.} 
\label{Fig:PL}
\end{figure}

The inset of Figure~\ref{Fig:PL} shows the normalized PL spectra. For comparison, the PL spectrum of a separate 1L WS$_2$ sample directly on quartz was also measured and is shown as the orange curve. The PL spectra exhibit similar lineshapes but systematic shifts in peak energy, which is plotted in the main panel. Going from quartz to bulk hBN, the PL peak redshifts from 2.0134 to 2.0023~eV. Under the bulk region of the hBN flake, with an estimated thickness of of 10 nm (see Figure S1), the graphene layer is not expected to affect the WS$_2$ layer. Therefore, this shift can be attributed to the enhanced dielectric screening by hBN, which has a larger dielectric constant than quartz. We note that dielectric screening reduces both the quasiparticle band gap and the exciton binding energy, with the two effects largely compensating each other.\cite{l115126802} Since the reduction of the band gap is slightly larger than that of the exciton binding energy, the net exciton energy decreases by 11.1~meV.

Going from bulk hBN to 6L hBN, the PL peak further redshifts to 2.0018~eV, indicating a small but measurable influence of the underlying graphene layer. Owing to the presence of mobile carriers, graphene is expected to provide stronger screening than hBN. Further decreasing the hBN thickness, and thus bringing graphene closer to WS$_2$, strengthens the screening effect, shifting the PL peak to 1.9951~eV for 1L hBN. When WS$_2$ is directly on graphene (denoted as 0L hBN), the redshift is the largest, with the PL peak located at 1.9899~eV.

These results confirm our hypothesis underlying the sample design, namely that the WS$_2$ and graphene 1Ls interact primarily through a proximity effect. Even a single layer of hBN is sufficient to prevent charge transfer and the formation of coupled electronic states between WS$_2$ and graphene. Therefore, this set of heterostructures provides an ideal platform for studying proximity effects in isolation from other interlayer interactions.

\subsection{Proximity Transient Absorption Spectra}

We first focus on the PTA in the heterostructure with 1L hBN, where the effect is expected to be the strongest. The sample was photoexcited by a pump pulse with a photon energy of 1.512 eV and a peak fluence of 360~$\upmu$J~cm$^{-2}$. Since this photon energy is well below the optical bandgap of WS$_2$, only the graphene layer is excited. The influence of these photoexcited carriers in graphene on the optical properties of WS$_2$ was probed using a time-delayed probe pulse with a tunable photon energy near the WS$_2$ exciton resonance at approximately 2.0~eV.

\begin{figure*}[t]
\centerline{\includegraphics[width=\textwidth]{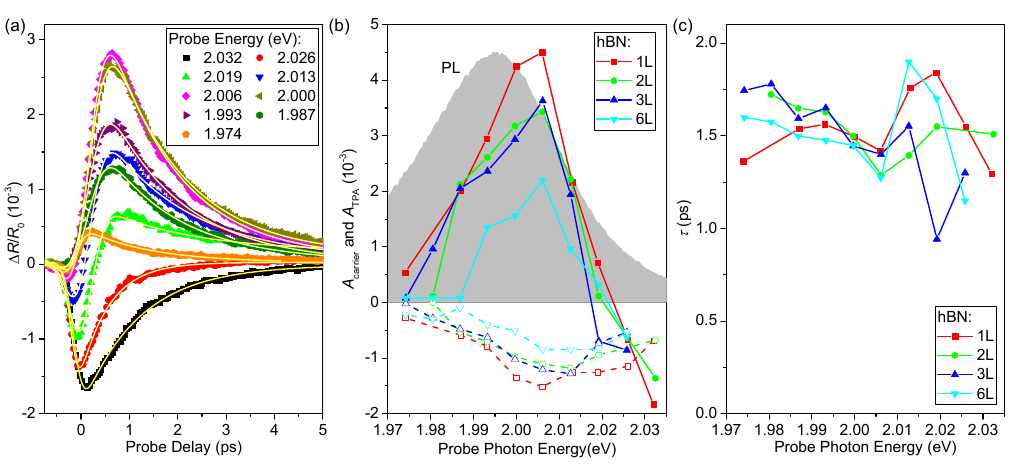}}
\caption{Proximity transient absorption spectroscopy. (a) Time-resolved differential reflectance of the Gr/1L-hBN/WS$_2$ heterostructure measured at various probe photon energies, as labeled. The yellow curves are fits to the data (see text). (b) Contributions from carrier-induced transient absorption ($A{_\mathrm{carrier}}$, solid red squares) and two-photon absorption ($A_{\mathrm{TPA}}$, open red squares) deduced from the fits shown in (a). The PL spectrum is replotted for comparison (shaded area). The values of $A_{\mathrm{carrier}}$ and $A_{\mathrm{TPA}}$ obtained from heterostructures with hBN thicknesses of 2L (green), 3L (blue), and 6L (cyan) are also shown. (c) Decay time constants deduced from fits of all four heterostructure samples.} 
\label{Fig:tas}
\end{figure*}

The symbols in Figure~\ref{Fig:tas}(a) show the time-resolved differential reflectance measured at various probe photon energies. The signal exhibits a strong dependence on probe energy within this spectral range, as well as rich temporal dynamics. We attribute these features to contributions to the signal from PTA due to the pump-injected carriers in graphene and from simultaneous absorption of pump and probe photons (two-photon absorption, TPA) when the two pulses overlap temporally:
\begin{equation}
S(t) = S_{\mathrm{carrier}}(t) + S_{\mathrm{TPA}}(t).
\end{equation}
The carrier density in graphene, $n(t)$, is governed by the rate equation taking into account both carrier injection and recombination processes,
\begin{equation}
\frac{dn(t)}{dt} = \frac{f_{\mathrm{abs}}\,I_{\mathrm{pump}}(t)} {E_{\mathrm{pump}}} - \frac{n(t)}{\tau},
\end{equation}
where $\tau$ is the carrier lifetime, $f_{\mathrm{abs}}$ is the absorbance of graphene,  $E_{\mathrm{pump}}$ is the pump photon energy, and the pump intensity $I_{\mathrm{pump}}(t)=I_0\exp[-t^2/(2\sigma_{\mathrm{pump}}^2)]$.

Since the probe pulse also has a finite width, this contribution to the signal is not directly proportional to $n(t)$. Instead, it is given by the convolution of the carrier density with the temporal profile of the probe pulse,
\begin{equation}
S_{\mathrm{carrier}}(t) \propto \int_{-\infty}^{\infty} n(t') I_{\mathrm{probe}}(t'-t)\,dt',
\end{equation}
where the probe pulse intensity $I_{\mathrm{probe}}(t'-t)=\exp[-(t'-t)^2/(2\sigma_{\mathrm{probe}}^2)]$. Using the associativity of convolution, the measured carrier response can be written as the convolution of an exponential decay function with the pump--probe cross-correlation function. Since both the pump and probe pulses are Gaussian, their cross-correlation is also Gaussian with width $\sigma_{\mathrm{c}}^{2} = \sigma_{\mathrm{pump}}^{2} + \sigma_{\mathrm{probe}}^{2}.$ The resulting carrier-induced transient signal becomes
\begin{equation}
S_{\mathrm{carrier}}(t) = A_{\mathrm{carrier}} \exp\left( -\frac{t}{\tau} \right) \left\{ 1+ \operatorname{erf} \left[ \frac{t-\sigma_{\mathrm{c}}^{2}/\tau} {\sqrt{2}\sigma_{\mathrm{c}}} \right] \right\},
\end{equation}
where $A_{\mathrm{carrier}}$ describes the magnitude of the carrier-induced signal and $\operatorname{erf}(x)$ is the error function defined as
\begin{equation}
\operatorname{erf}(x) = \frac{2}{\sqrt{\pi}} \int_{0}^{x} e^{-u^{2}}\,du.
\end{equation}

The second contribution due to TPA originates from simultaneous absorption of pump and probe photons by the sample. This nonlinear process increases the absorption of the probe pulse during its temporal overlap with the pump pulse, thereby reducing the reflected probe intensity and producing a negative differential reflectance signal. Since this process only occurs when the two pulses overlap temporally, its probe-delay dependence follows the pump--probe cross-correlation function:
\begin{equation}
S_{\mathrm{TPA}}(t) = A_{\mathrm{TPA}} \exp\left[ -\frac{t^{2}}{2\sigma^{2}} \right],
\end{equation}
where $A_{\mathrm{TPA}}$ is the amplitude of this contribution to the signal. By adding these two contributions, the signal can be well fit over the entire probe-energy range by Eq.~1, as shown by the curves in Figure~\ref{Fig:tas}(a). 

The red symbols in Figure~\ref{Fig:tas}(b) summarize the magnitude of these two components extracted from the fits. The magnitude of the carrier-induced differential reflectance, $A_{\mathrm{carrier}}$, exhibits a clear resonant feature near the excitonic resonance of WS$_2$, as indicated by the shaded area showing the replotted PL spectrum. This observation clearly demonstrates that the differential reflectance signal originates from the WS$_2$ layer. That is, photoexcited carriers in graphene induce a transient modification of the absorption spectrum of WS$_2$ through a proximity effect. The carrier-induced signal changes sign across the excitonic resonance, indicating a pronounced shift of the exciton transition energy caused by carrier-induced screening. This interpretation is consistent with the systematic PL shifts induced by hBN and graphene screening shown in Figure~\ref{Fig:PL}. In contrast, the TPA contribution remains negative over the entire probe-energy range, as expected since this process always causes additional absorption of the probe and thus reduced reflection. It also exhibits a relatively weak resonance near the exciton energy.

These measurements were repeated on heterostructure samples with hBN thicknesses of 2L, 3L, and 6L. The time-dependent differential reflectance data and their fits are shown in Figure S2. The extracted values of $A_{\mathrm{carrier}}$ and $A_{\mathrm{TPA}}$ are plotted in Figure~\ref{Fig:tas}(b). For each sample, $A_{\mathrm{carrier}}$ exhibits a clear resonance feature, with the magnitude decreasing overall with increasing hBN thickness. 

Finally, Figure~\ref{Fig:tas}(c) shows the carrier lifetimes extracted from the fits. The lifetime is essentially independent of both probe photon energy and hBN thickness. The observed time scale is similar to the carrier lifetime in an isolated graphene monolayer and is much shorter than the exciton lifetime in an isolated WS$_2$ monolayer, both measured using the same technique (to be discussed later). These observations further support the conclusion that the transient absorption signal at the WS$_2$ exciton resonance originates from photoexcited carriers in graphene through the proximity effect.

\subsection{Distance Dependence}

The measurements shown in Figure~\ref{Fig:tas} were performed by tuning the probe photon energy for each heterostructure sample, which inevitably requires minor realignment of the laser cavity and the measurement setup. Such adjustments can introduce small variations in the pulse width, spectral bandwidth, and time delay, potentially complicating quantitative comparisons among samples measured in different runs. To enable a more accurate comparison of samples with different hBN thicknesses, we fixed the probe photon energy at 2.010~eV, which yields the largest signal, and measured the time-dependent differential reflectance of all heterostructure regions of the sample under otherwise identical conditions. In this configuration, no realignment was required, ensuring consistent laser pulse characteristics throughout the measurements. The only change was the position of the laser spot on different regions of the sample.

\begin{figure}[t]
\centerline{\includegraphics[width=\columnwidth]{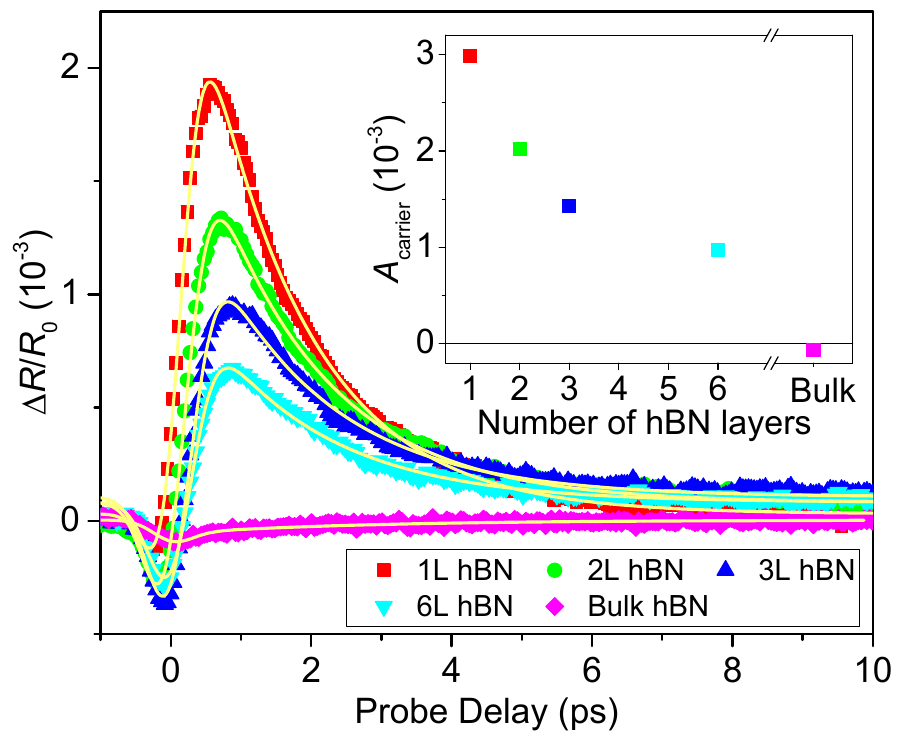}}
\caption{Time-resolved differential reflectance measured for samples with different hBN thicknesses, with pump and probe photon energies of 1.512 and 2.010~eV, respectively. The yellow curves are fits to the data using the model described in the text. The inset shows the amplitude of the carrier-induced transient absorption signal as a function of hBN thickness.} 
\label{Fig:thickness}
\end{figure}

The symbols in Figure~\ref{Fig:thickness} show the differential reflectance measured from regions with different hBN thicknesses, as labeled. As indicated by the yellow curves, all datasets can be fit well using the model discussed above. The extracted $A_{\mathrm{carrier}}$ is plotted as a function of hBN thickness in the inset. As the hBN thickness increases, the graphene layer becomes more separated from the WS$_2$ layer, reducing the influence of photoexcited graphene carriers on the excitonic properties of WS$_2$. Consequently, $A_{\mathrm{carrier}}$ decreases systematically with increasing hBN thickness. This behavior further confirms the origin of the observed signal as a proximity-induced modification of the WS$_2$ absorption spectrum by carriers in graphene. Interestingly, even for the 6L-hBN sample, corresponding to a graphene--WS$_2$ separation of approximately 2 nm, the PTA remains substantial, confirming the long-range nature of the excitonic Coulomb field in WS$_2$. With bulk hBN, the proximity effect is expected to be negligible. The small remaining negative signal can be attributed to the carrier-induced change in graphene absorption. This contribution should also be present in the other samples; however, it is largely masked by the much stronger (and positive) signal arising from the PTA.

\subsection{PTA for Noncontact Optical Modulation}

One potential application of the PTA is the modulation of the optical absorption of a 2D semiconductor in a noninvasive and noncontact manner. By introducing carriers into one layer (such as graphene) through optical excitation or electrostatic gating, the optical absorption of an adjacent 2D semiconductor (such as WS$_2$) can be tuned without directly injecting carriers into the semiconductor itself. To quantify the strength of this effect, we next investigate the dependence of the PTA on carrier density.

\begin{figure*}[t]
\centerline{\includegraphics[width=\textwidth]{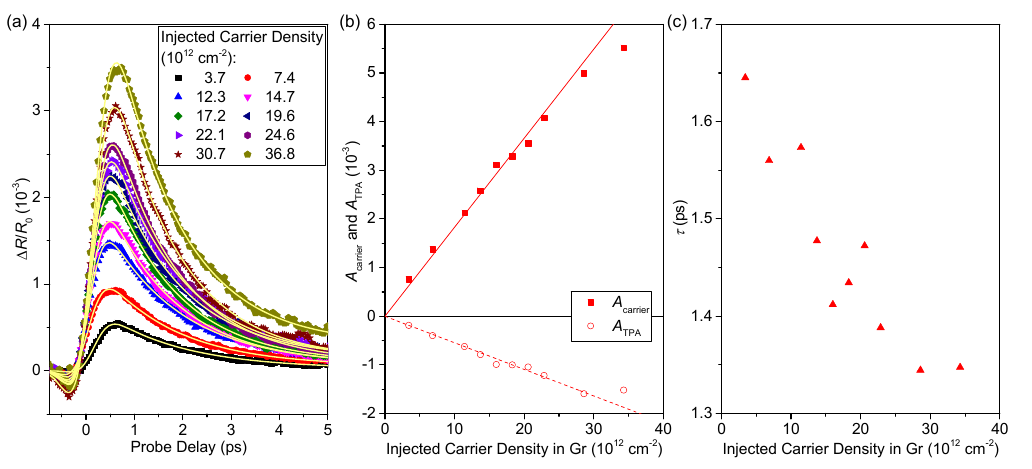}}
\caption{(a) Time-resolved differential reflectance of the Gr/1L-hBN/WS$_2$ heterostructure measured at different injected carrier densities, as labeled. The yellow curves are fits to the data using the model described in the text. (b) The extracted $A{_\mathrm{carrier}}$ (solid red squares) and $A_{\mathrm{TPA}}$ (open red circles) from the fits. The lines are linear fits. (c) The extracted carrier lifetime as a function of injected carrier density.} 
\label{Fig:power}
\end{figure*}

Figure~\ref{Fig:power}(a) shows the time-resolved differential reflectance of the Gr/1L-hBN/WS$_2$ sample measured at different pump fluences. Similar to the measurements presented in Figure~\ref{Fig:thickness}, these data were acquired without any tuning or realignment of the laser and pump-probe setup, ensuring accurate comparison among different fluences. The only adjustment was the rotation of a half-wave plate in the pump beam path before a polarizer, thereby varying the incident pump power on the sample.

To estimate the injected carrier density in graphene from the incident pump fluence, we used the sample reflectance of 6.9\% (measured) and a graphene absorbance of 1.5\%, which is a typical value for graphene on quartz. The resulting carrier densities are indicated in the label of Figure~\ref{Fig:power}(a). As shown by the yellow curves, the data can be well fit using the model described above. We find that $A_{\mathrm{carrier}}$ scales linearly with the carrier density [solid squares in Figure~\ref{Fig:power}(b)]. We define a PTA cross section, $\sigma_{\mathrm{PTA}}$, as the differential reflectance induced per unit carrier density, $A_{\mathrm{carrier}}=\sigma_{\mathrm{PTA}} n$. From the linear fit shown in Figure~\ref{Fig:power}(b), we obtain $\sigma_{\mathrm{PTA}} = 1.8 \times 10^{-16}~\mathrm{cm}^2$. Similarly, the TPA contribution (open circles and dashed line) also scales linearly with pump fluence, since this process involves one probe photon for each pump photon participating in the nonlinear interaction. Finally, the extracted carrier lifetime, shown in Figure~\ref{Fig:power}(c), decreases slightly with increasing carrier density. This behavior is characteristic of exciton--exciton annihilation in 2D semiconductors,\cite{b89125427,nl145625,b93201111,l124057403} where enhanced exciton--exciton interactions at higher excitation densities introduce an additional nonradiative recombination channel and accelerate the decay dynamics.

The measured PTA cross section allows us to evaluate the practical potential of this effect for optical modulation. With 1L hBN, a reflectance change of 1\% of WS$_2$ would require a carrier density in the graphene layer of approximately $6 \times 10^{13}$~cm$^{-2}$. Although this density is higher than those used in the present experiments, it is within the range accessible by electrostatic control. Conventional dielectric gating can routinely induce carrier densities approaching $10^{13}$~cm$^{-2}$ in graphene, while electrolyte and ionic-liquid gating have demonstrated densities exceeding $10^{14}$~cm$^{-2}$.\cite{pnas10813002}  For comparison, we also performed differential reflectance measurements on a WS$_2$ monolayer on quartz (Figure S3). From these measurements, we obtained a modulation cross section of approximately $6.7\times10^{-16}~\mathrm{cm}^2$. Thus, the modulation efficiency of PTA is only 3.7 times smaller than that achieved by directly injecting carriers into the WS$_2$ layer. 

Despite the slightly smaller cross section, the PTA offers an attractive approach for optical modulation for two reasons. First, it provides a noninvasive means of controlling the optical properties of WS$_2$ without introducing carriers directly into the semiconductor. This feature is particularly advantageous for applications where preserving the intrinsic electronic and excitonic properties of the active layer is desirable. Moreover, the effect remains substantial even when the graphene and WS$_2$ layers are separated by several atomic layers. For example, with a 6L hBN spacer, the modulation amplitude is still approximately one-third of that obtained with a 1L hBN spacer [Figure~\ref{Fig:tas}(b)]. This robustness allows the effect to be utilized in more complex multilayer van der Waals heterostructures while maintaining electrical and chemical isolation between functional layers. Second, the modulation bandwidth is potentially much higher than that of conventional excitonic modulation in TMDs. The carrier lifetime in graphene is less than 2 ps, corresponding to a characteristic bandwidth approaching 0.5 THz. In contrast, the exciton population in monolayer WS$_2$ persists for approximately 100 ps (Figure S3), limiting the modulation bandwidth to roughly 10 GHz. The contrast becomes even more pronounced in TMD heterostructures, where interlayer exciton lifetimes can reach hundreds of nanoseconds. Such long-lived excitations can severely restrict device speed. Hence, PTA offers a promising route toward high-speed optical modulation while retaining the strong optical response of TMD materials.

\subsection{PTA for Remote Probing of Carrier Dynamics}

Beyond optical modulation, PTA also provides a new approach for probing carrier dynamics in materials that are difficult to study using conventional optical techniques. Conventional transient absorption spectroscopy relies on detecting photoinduced changes in optical transitions of the material under investigation. As a result, its sensitivity relies on the strength and accessibility of those optical transitions. For many materials, the resulting transient signals can be weak or difficult to detect, motivating the development of alternative probing schemes.

\begin{figure}[t]
\centerline{\includegraphics[width=\columnwidth]{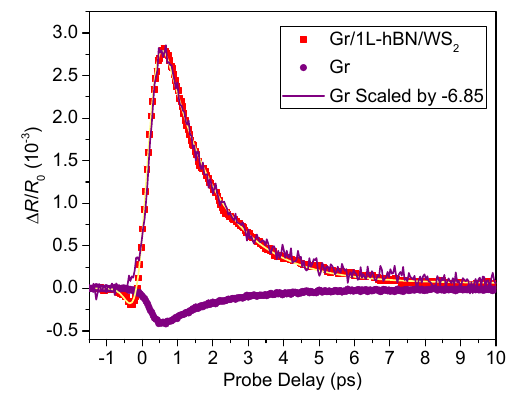}}
\caption{Time-resolved differential reflectance of Gr/1L-hBN/WS$_2$ (red squares) and Gr (purple circles) under identical conditions. The yellow curve is a fit. The purple curve is scaled from the purple circles by a factor of $-6.85$.} 
\label{Fig:probe}
\end{figure}

To illustrate the potential of PTA as such an alternative, we compare the transient optical response of graphene measured directly with that detected through a nearby WS$_2$ monolayer. The purple circles in Figure~\ref{Fig:probe} show the differential reflectance measured from a region of bare graphene on quartz using a 1.512~eV pump pulse with a fluence of 360~$\upmu$J~cm$^{-2}$ and a 2.010~eV probe pulse. The negative signal is consistent with photoinduced absorption by pump-generated carriers, which absorb probe photons and thereby reduce the reflected intensity. When the same measurement is performed on the Gr/1L-hBN/WS$_2$ region, by simply moving that region into the laser spots, a substantially larger signal is observed (red squares). As discussed above, the data are well described by the model (yellow curve), yielding a carrier decay time of approximately 1.5~ps. The purple curve is obtained by multiplying the graphene signal (purple circles) by a factor of $-6.85$. The nearly perfect overlap between the purple curve and the PTA signal demonstrates that the latter faithfully probes the carrier dynamics in graphene while providing an enhancement in sensitivity of nearly seven times.

These results suggest a general strategy for probing carrier dynamics in materials that are not readily accessible by direct optical measurements. By placing the material of interest in close proximity to a sensing layer (WS$_2$ monolayer or another material with strong optical coupling), separated by a thin insulating spacer, carriers in the target material can modulate the optical absorption of the sensing layer through the screening effect. The sensing layer thus acts as an optical transducer that converts the carrier population in the neighboring material into a differential reflectance signal. If the sensing layer (such as a TMD monolayer) possesses strong excitonic resonances and large optical cross sections, the resulting signal can be substantially stronger than that obtained by probing the target material directly. Although graphene can still be measured optically in this proof-of-concept measurement, the same approach could be particularly valuable for materials with indirect or wide bandgaps, materials whose relevant optical transitions lie outside the spectral range accessible to available lasers, or systems that interact only weakly with light.

\subsection{Interlayer Coupling between Graphene and WS$_2$}

The Gr/hBN/WS$_2$ heterostructures discussed above establish that graphene carriers can modulate the exciton absorption of a nearby WS$_2$ layer through PTA. An important question is whether the same mechanism remains valid when the hBN spacer is removed and graphene is brought into direct contact with WS$_2$. If direct electronic coupling were negligible, the Gr/WS$_2$ heterostructure would be expected to follow the trend established by the hBN-separated samples, with the signal amplitude increasing as the interlayer separation approaches zero while retaining similar spectral characteristics and carrier dynamics. However, if direct graphene--WS$_2$ coupling becomes important, qualitatively different spectral features and dynamics may emerge.

Previous studies have established substantial interlayer coupling in directly contacted graphene--WS$_2$ heterostructures. For example, time- and angle-resolved photoemission measurements directly observed ultrafast charge transfer following resonant excitation of WS$_2$, resulting in a transient charge-transfer state.\cite{sa6eaay0761} Combined with theoretical calculations, these studies showed that charge transfer is mediated by hybridized interlayer states arising from coupling between graphene and WS$_2$ electronic bands.\cite{2dm10035025} The transfer dynamics depend sensitively on excitation energy, indicating the participation of different hybridized states in distinct transfer pathways.\cite{nl251214} In addition, proximity-induced spin--orbit coupling has been observed in graphene/WS$_2$ heterostructures, providing further evidence that direct contact with WS$_2$ substantially modifies the electronic states of graphene.\cite{l120106802} Similar behavior has been reported in other graphene--TMD systems.\cite{acsnano1111678,nl218051} Together, these studies indicate that direct graphene--TMD contact can support coupled electronic states, consistent with theoretical predictions.\cite{b93155104,l119146401}

\begin{figure*}[t]
\centerline{\includegraphics[width=\textwidth]{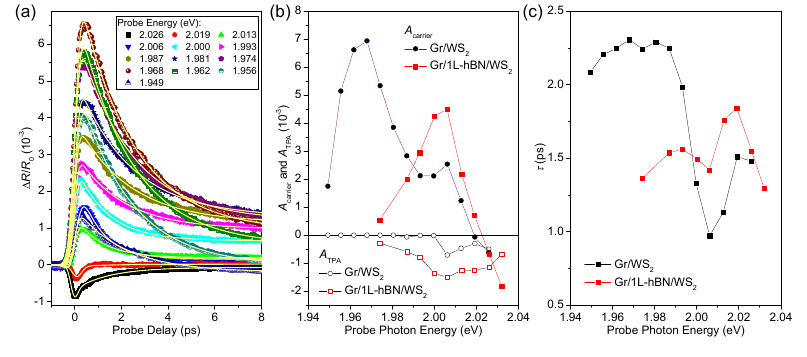}}
\caption{Transient absorption spectroscopy of the Gr/WS$_2$ heterostructure. (a) Time-resolved differential reflectance measured at various probe photon energies, as labeled. The yellow curves are fits to the data (see text). (b) Carrier-induced differential reflectance amplitude, $A_{\mathrm{carrier}}$ (solid purple circles), and two-photon absorption amplitude, $A_{\mathrm{TPA}}$ (open purple circles), extracted from the fits in (a). Results for Gr/1L-hBN/WS$_2$ from Figure~\ref{Fig:tas} are shown for comparison (red squares). (c) Carrier decay time constants extracted from the fits in (a).}
\label{Fig:GrWS2}
\end{figure*}

In our Gr/hBN/WS$_2$ heterostructure samples, the hBN spacer, even in the case of 1L, suppresses charge transfer and wavefunction overlap between graphene and WS$_2$, such that the two materials remain electronically isolated. This interpretation is supported by the observation that the transient signal follows the carrier dynamics of graphene while showing a resonance near the A-exciton transition of WS$_2$. The optical response therefore originates from modulation of the WS$_2$ exciton absorption by graphene carriers through Coulomb screening rather than from the formation of coupled electronic states. As the hBN thickness increases from 1L to 6L, the signal amplitude decreases monotonically, while the spectral shape and carrier dynamics remain essentially unchanged, consistent with a progressive weakening of the screening interaction with increasing separation.

To determine whether this behavior persists in the absence of the spacer, we performed transient absorption measurements on the Gr/WS$_2$ heterostructure (labeled as 0L in Figure~\ref{Fig:PL}) using a 1.512~eV pump pulse with a fluence of 360~$\upmu$J~cm$^{-2}$. Figure~\ref{Fig:GrWS2}(a) shows the time-resolved differential reflectance with various probe photon energies. As with the hBN-separated samples, the data can be fit well by the model discussed above, allowing the carrier- and TPA-induced differential reflectance amplitudes, as well as the carrier decay time constants, to be extracted, as summarized in Figures~\ref{Fig:GrWS2}(b) and (c).

The extracted transient absorption spectra and carrier dynamics reveal behavior that differs qualitatively from that of the Gr/hBN/WS$_2$ heterostructures. While the hBN-separated samples exhibit a single resonance near 2.0~eV, the directly contacted Gr/WS$_2$ heterostructure displays two distinct resonances. One remains near 2.0~eV and closely resembles the PTA feature observed in the hBN-separated samples. A second resonance emerges at approximately 1.96~eV and reaches an amplitude more than twice that of the 2.0~eV feature. Because the 1.96~eV feature is absent in all samples containing an hBN spacer and is substantially below the 1.99~eV PL peak of this sample, it is naturally associated with direct interaction between graphene and WS$_2$. Additional evidence is provided by the carrier dynamics. The lower-energy resonance is characterized by a decay time of approximately 2.2~ps, as shown in Figure~\ref{Fig:GrWS2}(c), whereas the resonance near 2.0~eV exhibits a shorter lifetime of about 1.5~ps, similar to that observed in the hBN-separated samples. The coexistence of two resonances with different amplitudes and distinct temporal dynamics indicates that they originate from different photoexcited states or relaxation pathways. We therefore attribute the 1.96~eV resonance to an interfacial optical response associated with direct graphene--WS$_2$ coupling, whereas the resonance near 2.0~eV is attributed to the PTA mechanism discussed above.

The appearance of the additional 1.96~eV resonance therefore indicates that the zero-separation limit is not fully described by a simple extrapolation of PTA. Instead, direct graphene--WS$_2$ contact introduces additional optical responses associated with interlayer coupling. These observations indicate the importance of the hBN spacer for more clearly identifying PTA. In the absence of hBN, direct-contact vdW heterostructures often exhibit modified optical and electronic properties arising from interlayer interactions, making it difficult to distinguish purely electrostatic proximity effects from other interfacial phenomena. In our previous studies of Gr/WS$_2$\cite{nc55622} and Gr/MoS$_2$/MoSe$_2$/Gr heterostructures,\cite{acsnano1725368} we observed transient optical features demonstrating that photoexcited carriers in graphene can influence the absorption of a neighboring TMD layer. However, those studies could not differentiate effects of PTA or direct coupling. Furthermore, Li \textit{et al.}\cite{acsnano155495} reported transient bleaching of a WS$_2$ resonance induced by excitons in an adjacent MoTe$_2$ layer in a type-I MoTe$_2$/WS$_2$ heterostructure and attributed the effect to modulation of the WS$_2$ exciton oscillator strength by the electric field of MoTe$_2$ excitons. Nevertheless, the two layers remain in direct contact, and a redshift of the WS$_2$ exciton resonance was observed and attributed in part to interlayer electronic coupling. Therefore, although the proposed proximity mechanism is plausible and supported by various features observed, contributions from interfacial electronic states and other consequences of direct electronic coupling cannot be completely excluded. In the present study, the insertion of an hBN spacer with various thicknesses diminished direct electronic coupling while preserving long-range electrostatic interactions, allowing PTA to be isolated from direct-contact interfacial effects.

\section{Conclusions}

In summary, we have demonstrated a proximity transient absorption effect in Gr/hBN/WS$_2$ van der Waals heterostructures. By selectively photoexciting carriers in graphene and probing the excitonic absorption of a nearby WS$_2$ monolayer, we observed a transient optical response that originates from modulation of the WS$_2$ exciton by graphene carriers through Coulomb screening. The signal exhibits a clear excitonic resonance, follows the carrier dynamics of graphene, and decreases systematically with increasing hBN thickness, confirming its proximity-induced origin. Remarkably, the effect remains substantial even when graphene and WS$_2$ are separated by several atomic layers, highlighting the long-range nature of the interaction.

The demonstrated PTA provides a noninvasive means of controlling and probing the optical properties of 2D semiconductors. The measured modulation cross section indicates that optical absorption in a TMD monolayer can be tuned without directly introducing carriers into the semiconductor. In addition, by using the TMD exciton resonance as an optical transducer, carrier dynamics in a neighboring material can be detected with substantially enhanced sensitivity compared with direct optical probing. These features make PTA a promising approach for investigating materials that are difficult to study using conventional transient absorption spectroscopy.

More broadly, these results have important implications for the interpretation of transient absorption measurements in van der Waals heterostructures. Such experiments are commonly analyzed under the assumption that a transient signal observed at the optical resonance of a given layer directly reflects the dynamics of photoexcited carriers or excitons within that layer. The present work demonstrates that this assumption is not always valid. Proximity effects should be carefully considered when interpreting transient optical measurements of van der Waals heterostructures, particularly in systems where strong Coulomb interactions exist between electronically isolated layers. Furthermore, while the Gr/hBN/WS$_2$ heterostructures exhibit behavior consistent with electronically isolated layers interacting only through Coulomb screening, the directly contacted Gr/WS$_2$ heterostructure displays an additional resonance with distinct temporal dynamics. This feature is consistent with the formation of interfacial states previously reported in directly contacted graphene--TMD heterostructures.

Together, these results establish PTA as a new tool for studying interlayer interactions in van der Waals heterostructures and suggest new opportunities for optical modulation, carrier sensing, and the investigation of coupled electronic states at atomically thin interfaces.

\section{Experimental Section}

Transient absorption measurements were carried out at room temperature in ambient atmosphere using a home-built microscope-integrated pump–probe system. The ultrafast light source comprises a Ti:sapphire oscillator (Newport Mai Tai) coupled to an optical parametric oscillator (Newport Inspire HF). The Mai Tai generates 100 fs pulses centered at 820 nm with a repetition rate of 80 MHz. These pulses are frequency doubled within the Inspire to produce 410 nm radiation, which drives the parametric process. The tunable signal output serves as the probe beam. The residual 820 nm beam is focused onto another nonlinear optical crystal to produce its second harmonic at 410 nm, employed as the pump. The pump and probe beams are spatially overlapped using a beamsplitter and focused onto the sample through a microscope objective, yielding beam diameters of approximately 1–2~$\upmu$m. The reflected probe light is collected by the same objective and detected with a silicon photodiode. To eliminate contributions from scattered and reflected pump light, appropriate optical filters are inserted before the photodiode. The photodiode signal is analyzed using a lock-in amplifier (Stanford Research Systems SR830) referenced to a mechanical chopper modulating the pump beam at approximately 1 kHz. The experiment measures the differential reflectance, $\Delta R/R_0$, where $R$ and $R_0$ denote the sample reflectance in the presence and absence of the pump pulse, respectively. Time-resolved measurements are obtained by varying the delay between the pump and probe pulses. This is accomplished by changing the optical path length of the pump beam with a retroreflector mounted on a motorized translation stage.

\begin{acknowledgments}

Research supported by the U.S. Department of Energy, Office of Basic Energy Sciences, Division of Materials Sciences and Engineering under Award DE-SC0020995.
\end{acknowledgments}

%

\clearpage
\onecolumngrid

\begin{center}
{\bf Supporting Figures}
\end{center}

\setcounter{figure}{0}
\renewcommand{\figurename}{Figure}
\renewcommand{\thefigure}{S\arabic{figure}}

\begin{figure*}[h]
\centerline{\includegraphics[width=8.5cm]{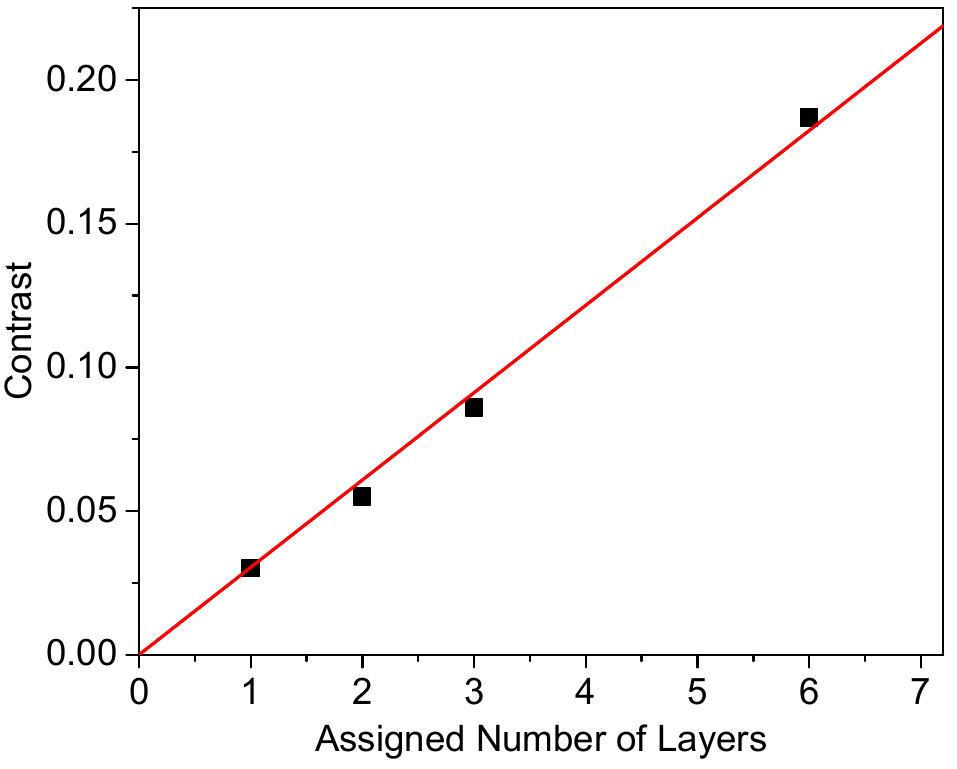}}
\caption{Optical contrast method to determine hBN thickness. Figure~2(b) shows the optical microscopy image of the hBN flake on PDMS, which was used to fabricate the heterostructures. Since the PDMS substrate is transparent and much thicker than the focal depth of the microscope objective, the contribution from the back surface of the PDMS can be neglected. The image contrast is therefore mainly determined by the interference between the light reflected from the air/hBN and hBN/PDMS interfaces. In the few-layer limit, the optical phase accumulated across the hBN layer is much smaller than $2\pi$, so the phase difference between the two reflection fields is approximately proportional to the hBN thickness. Consequently, the optical contrast, defined as $(I_{\mathrm{flake}}-I_{\mathrm{PDMS}})/I_{\mathrm{PDMS}}$, is expected to vary linearly with layer number. As shown by the black symbols, the observed discrete green-channel contrasts show clear step-like behavior, with the smallest contrast being 0.0301. This step-like linear dependence allows us to assign the number of hBN layers of different regions. The region labeled as bulk exhibits a green-channel contrast of 0.881. Using the linear relation established from the few-layer regions, this contrast corresponds to approximately 30 hBN layers, or a thickness on the order of 10 nm.} 
\label{Fig:SIcontrast}
\end{figure*}

\begin{figure*}[h]
\centerline{\includegraphics[width=16.5cm]{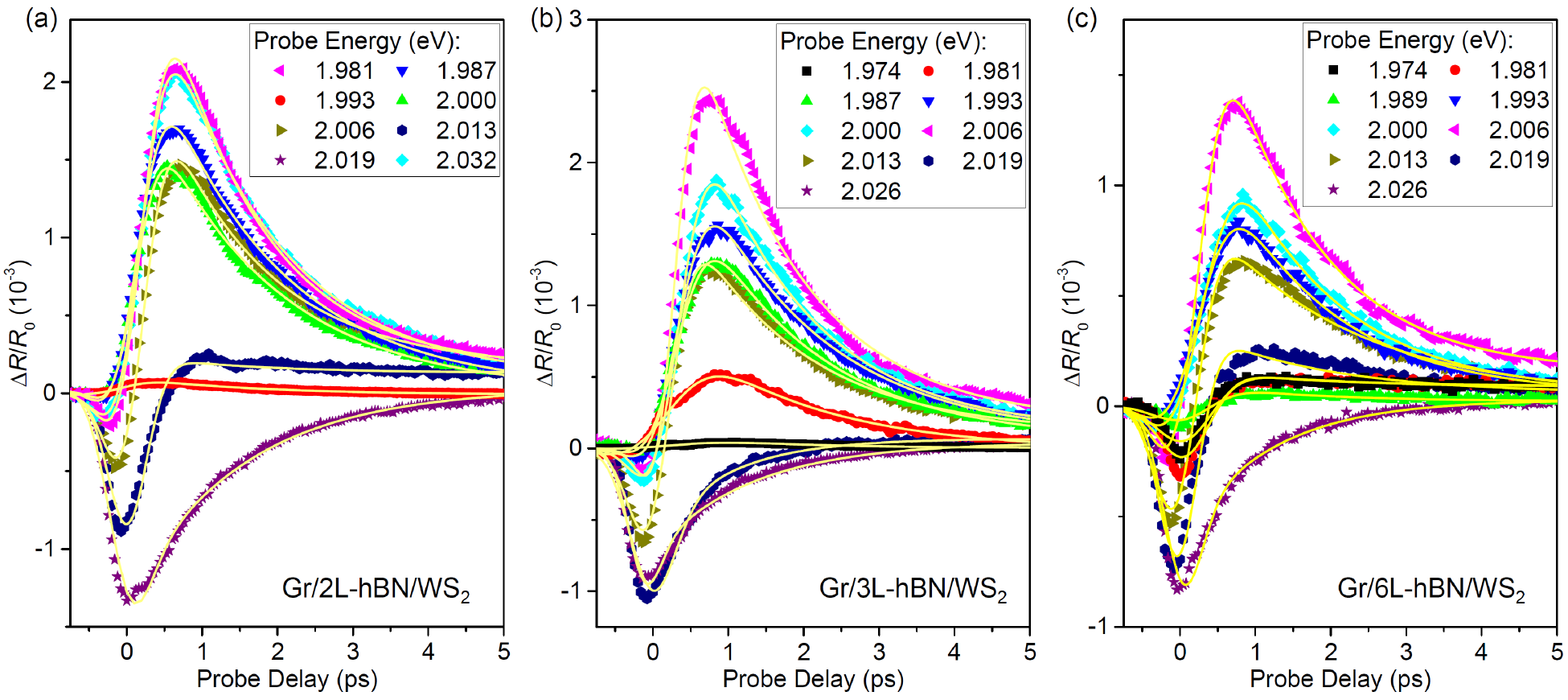}}
\caption{Time-resolved differential reflectance of the Gr/hBN/WS$_2$ heterostructures with hBN thicknesses of 2L, 3L, and 6L, shown in panels (a), (b), and (c), respectively. All measurements were performed using a 1.512 eV pump pulse with a fluence of 360~$\upmu$J~cm$^{-2}$, identical to the conditions used for the data shown in Figure~4(a) of the main text. The yellow curves are fits to the data using the model described in the main text. The extracted parameters are summarized in Figure~4(b) and 4(c) of the main text.} 
\label{Fig:SItas236corp}
\end{figure*}

\begin{figure*}[h]
\centerline{\includegraphics[width=16.5cm]{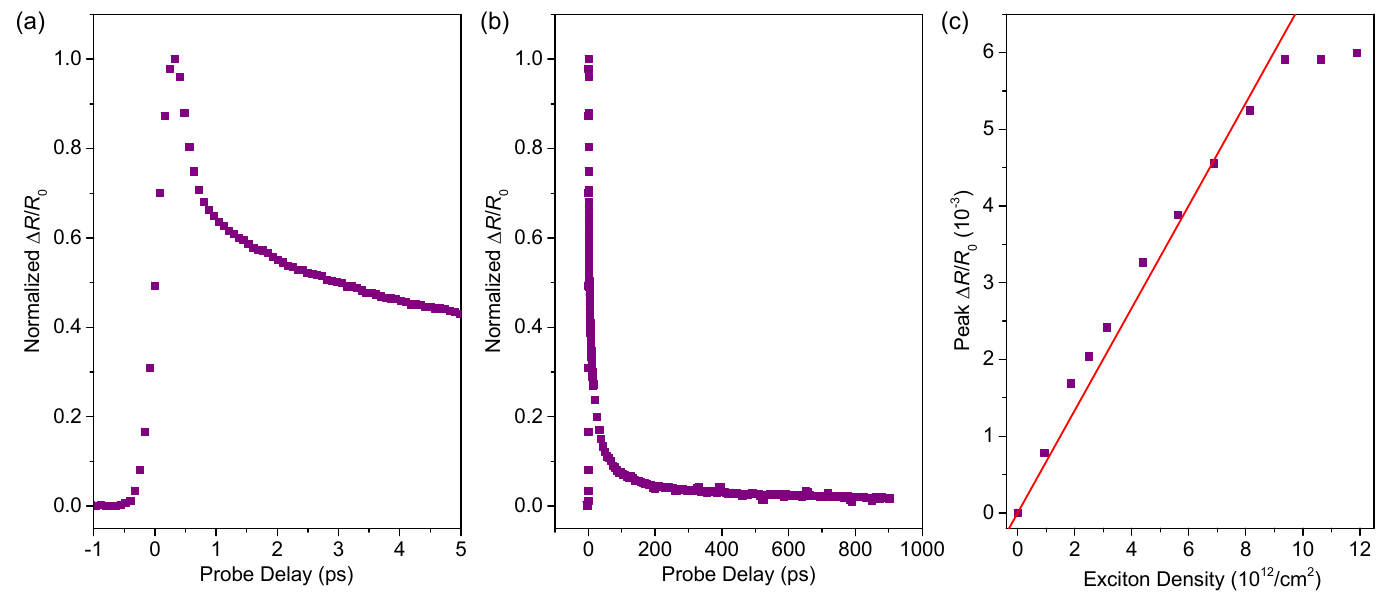}}
\caption{(a) Differential reflectance of a 1L WS$_2$ on a quartz substrate measured with a 3.024 eV pump and a 2.006 eV probe. (b) Same set of data over a long time range. (c) Peak differential reflectance as a function of the injected exciton density. The exciton densities were estimated from the pump fluence by using a reflectance of 6.9\% and an absorbance of 0.12. The linear fit (red line) indicates a cross section of $6.7 \times 10^{-16}$~cm$^2$.} 
\label{Fig:SIWS2}
\end{figure*}

\end{document}